# HHG-laser-based time- and angle-resolved photoemission spectroscopy of quantum materials

Takeshi Suzuki[1*], Shik Shin[2,3], and Kozo Okazaki[1,3,4,†]

[1]*The Institute for Solid State Physics, The University of Tokyo, Kashiwa, Chiba 277-8581 Japan*
[2]*Office of University Professor, The University of Tokyo, Kashiwa, Chiba 277-8581 Japan*
[3]*Material Innovation Research Center, The University of Tokyo, Kashiwa, Chiba 277-8561, Japan*
[4]*Trans-Scale Quantum Science Institute, The University of Tokyo, Bunkyo-ku, Tokyo 113-0033, Japan*

Abstract
Time- and angle-resolved photoemission spectroscopy has played an important role in revealing the non-equilibrium electronic structures of solid-state materials. The implementation of high harmonic generation to obtain a higher photon energy also allows us to investigate the wide Brillouin zone on a time scale below 100 fs. In this article, we review our recent studies using high-harmonic-generation-laser-based time- and angle-resolved photoemission spectroscopy to study a variety of quantum materials. We reveal many unprecedented phenomena in each system and highlight some representative results.

## 1 Introduction

The electronic band structure is one of the most fundamental aspects of a material. By applying the photoelectric effect, photoemission spectroscopy can directly observe the electronic band structure of a material and has served as an extremely powerful experimental method for decades [1]. The implementation of laser to photoemission spectroscopy has dramatically boosted its power by highlighting unique advantages of laser.

By using the monochromaticity of the laser, the energy resolution of photoemission spectroscopy has been significantly improved, which is far beyond the improvements achieved by the development of synchrotron radiation facilities and electron analyzers [2][3], and enabled us to observe fine structures [4].

The application of the pulsed nature of laser enables measurements in a time-resolved manner [5][6]. In particular, the mode-locking and amplification techniques using a Ti:Sapphire crystal as a gain medium enabled sufficient photon flux to be achieved within an extreme ultraviolet wavelength region by using wavelength conversion techniques, and the time resolution can be achieved at a femtosecond time scale [7][8][9][10][11]. However, photon energies of ~6 eV are more commonly used as a light source for time- and angle-resolved photoemission spectroscopy (TARPES) through the use of up-conversion techniques with a nonlinear crystal such as $\beta$-$BaB_2O_4$(BBO), and the energy and momentum regions accessible with these photon energies are extremely limited [12][13][14].

Alternatively, high harmonic generation (HHG) techniques using noble gas overcome this limitation, and can generate much higher-order harmonics within the energy region of 10–70 eV, and enable access to full valence bands, shallow core levels, and a wide momentum space [15][16][17][18][19][20][21][22][23][24][25].

In this review article, we briefly present our recent results measured using HHG-laser-based time- and angle-resolved photoemission spectroscopy (HHG laser TARPES) [26][27][28][29][30][31]. First, we briefly describe our experimental setup followed by the recent results of representative quantum materials, which consist of iron-based superconductors, graphene, and excitonic insulators, and we conclude with an outlook on the future progress to be made in this field.

## 2 Experimental setup

A schematic illustration of the HHG laser TARPES system is shown in Fig. 1. We used two types of Ti:sapphire amplification systems with different repetition rates of 1 kHz (Coherent, Astrella) and 10 kHz (Spectra Physics, Solstice Ace). For both systems, the center wavelength was 800 nm, and the time duration was 35 fs. The pulse energy is 6 mJ for 1 kHz and 0.7 mJ for 10 kHz, respectively. The advantage of using 1 kHz is to achieve higher pump excitation while that of 10 kHz is to achieve better signal-to-noise ratio by reducing space charge effects with keeping higher photoemission count rate. We used the 1 kHz system in the sections 3.1.1., 3.1.2., 3.3.1 and used 10 kHz system in the other sections. The fundamental beam was split between the pump and probe beams. For the pump pulses, we use the fundamental wavelength in this review article and can change the fluence using a half-wave plate and a polarizer. For the probe pulses, we first double the photon energy to 3.10 eV using a BBO crystal, and then focus the pulses on Ar gas filled in the gas cell to achieve HHG. We typically select the 9(7)th harmonic corresponding to 27.9(21.7) eV by using a pair of SiC/Mg multilayer mirrors for a 1(10) kHz system. Photoelectrons were collected using a hemispherical electron analyzer (Scienta Omicron, R4000). A typical time resolution of ~70 fs was obtained by measuring the response time of highly oriented pyrolytic graphite as a reference sample. The energy resolution was set to 250 meV. The base pressure of the analyzer chamber was ~$2 \times 10^{-11}$ Torr.

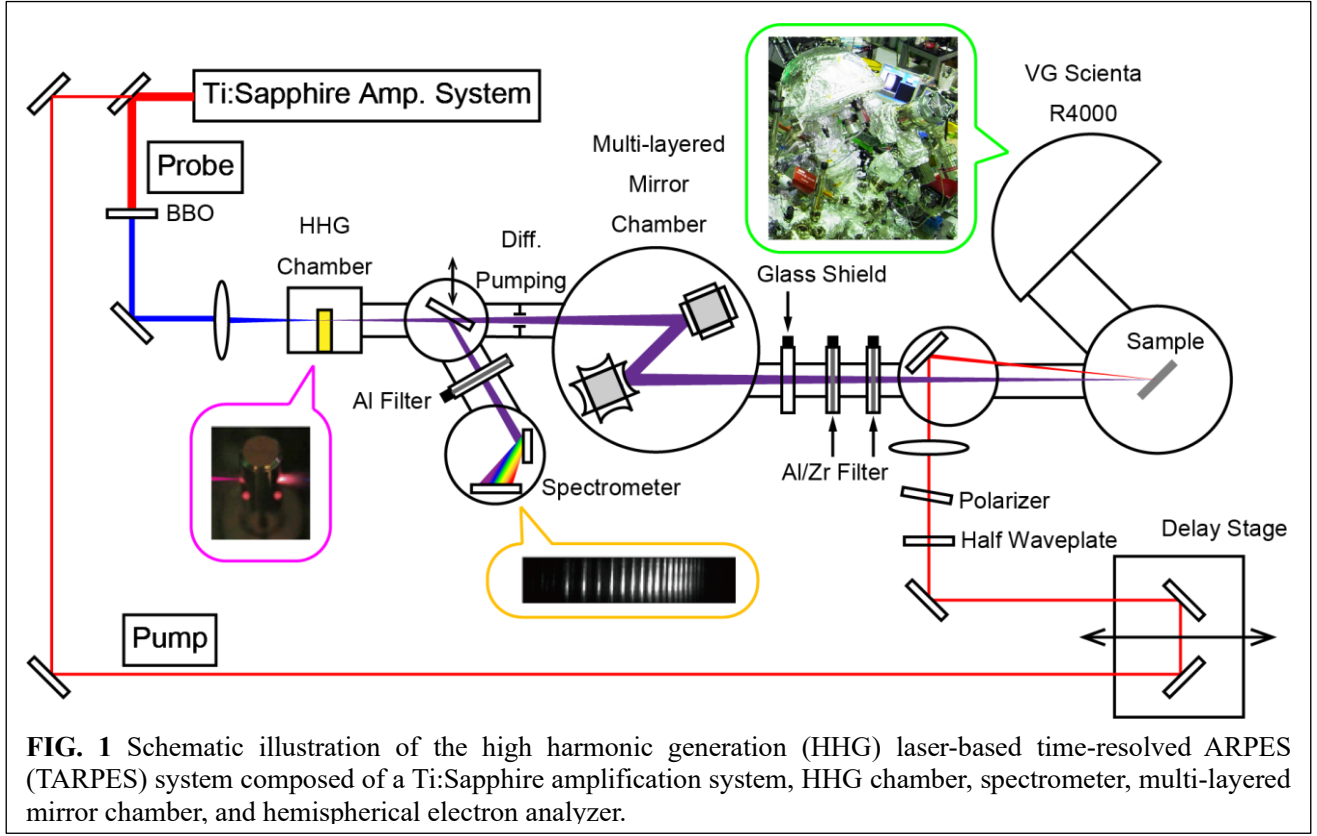

**FIG. 1** Schematic illustration of the high harmonic generation (HHG) laser-based time-resolved ARPES (TARPES) system composed of a Ti:Sapphire amplification system, HHG chamber, spectrometer, multi-layered mirror chamber, and hemispherical electron analyzer.

## 3 Results
### 3.1 Fe-based superconductors

Iron-based superconductors exhibit the second highest critical temperature ($T_c$) at ambient pressure following cuprate superconductors. In addition to the unsettled superconducting mechanisms [4] [32] [33] [34] [35] [36], they provide rich and exotic physical properties, including electronic nematicity [37] [38] [39], a Bardeen-Cooper-Shriefer (BCS) to Bose-Einstein condensation (BEC) crossover [40] [41] [42] [42], or the emergence of Majorana fermions on topological superconducting surfaces [43] [44]. The nonequilibrium profiles of iron-based superconductors have also been studied using numerous methods. The reported phenomena range from a photoinduced chemical potential shift [45], the generation of spin-density waves [46], and excitonic states [47].

We have also used HHG laser TARPES to study iron-based superconductors, namely, $BaFe_2As_2$ [26] and FeSe [27]. In both cases, we found significant modulations of the Fermi surfaces as a result of the generation of coherent phonons. From the calculation results based on density functional theory (DFT), we found that the observed modulations were ascribed to the modulation of the lattice structure, which might be associated with photoinduced superconductivity.

#### 3.1.1 $BaFe_2As_2$

Whereas the parent compound, $BaFe_2As_2$, does not show superconductivity at ambient pressure, superconductivity can be induced under various conditions such as hole doping through the substitution of K for Ba [48], electron doping through the substitution of Co for Fe [49], and the isovalent substitution of P for As [50], as well as under high pressure [51]. These intriguing properties prompted us to search for the emergence of superconductivity through photoexcitation.

Figures 2(a) and 2(b) show the momentum-integrated TARPES spectra across the hole and electron Fermi surfaces (FSs) as a function of energy with respect to the Fermi level ($E_F$) and pump-probe delay time. For both FSs, the spectra show that electrons are immediately excited upon photoexcitation, followed by relatively slow relaxation dynamics. In addition, it was noted that the oscillatory components were superimposed onto the overall background electron dynamics. These features are more clearly recognized in Figs. 2(c) and 2(d), where the integrated intensities above $E_F$ corresponding to the boxed regions in Figs. 2(a) and 2(b) are shown, respectively. To highlight the oscillatory components, we subtracted the background signal, denoted by the dashed lines in Figs. 2(c) and 2(d), fitted by an

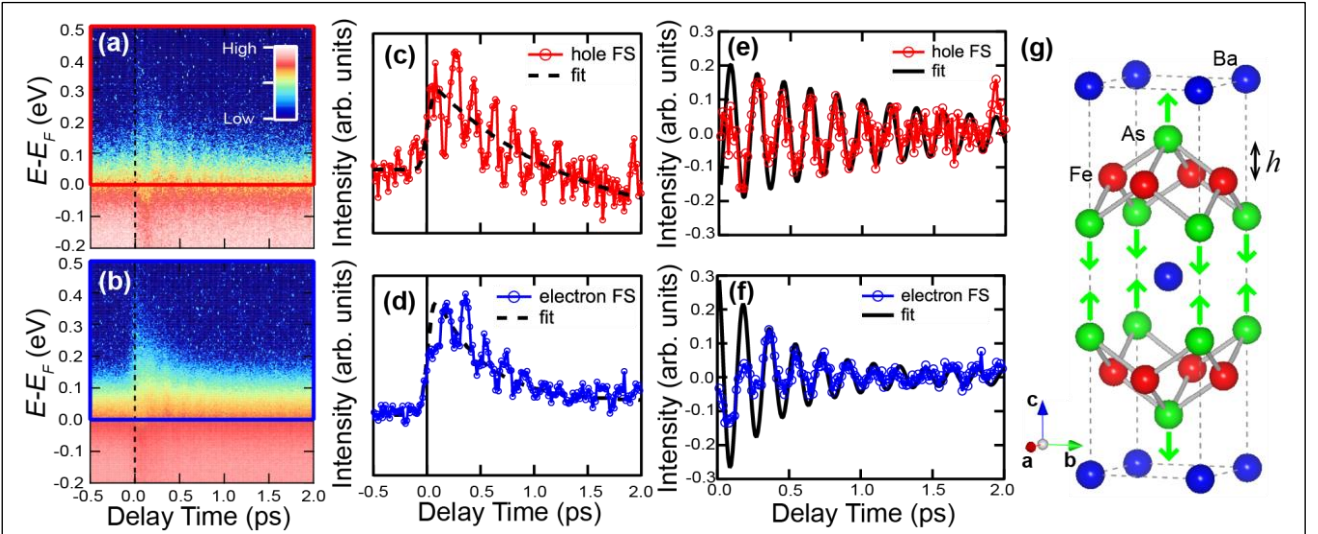

**FIG. 2 (a)**, **(b)** Momentum-integrated TARPES spectra across the hole and electron FSs as a function of energy ($E$) with respect to $E_F$ and pump-probe delay time. **(c)**, **(d)** Integrated intensity corresponding to the regions surrounded by the red and blue boxes in Figs. 2(a) and 2(b), respectively. **(e), (f)** Oscillatory components of the hole and electron FSs, which are obtained by subtracting the carrier dynamics from **(c)** and **(d)**, respectively. **(g)** Crystal structure of BaFe$_2$As$_2$ and the definition of the pnictogen height $h$. Thick arrows indicate the displacement of the As atoms corresponding to the $A_{1g}$ phonon.

exponential decay function plus a residual slowly decaying component convoluted using a Gaussian. The oscillatory components of the hole and electron FSs are shown in Figs. 2(e) and 2(f), respectively, with the fits of the damped oscillation functions. It can be clearly seen that the hole and electron FSs exhibit antiphase oscillations with respect to each other. The frequency for both oscillations is found to be 5.5 THz, which corresponds to the $A_{1g}$ phonon mode shown in Fig. 2(g). From the fitting analyses, both oscillations show cosine-like behaviors that are a signature of the displacive excitation of coherent phonons (DECPs). According to the DECP mechanism, it is considered that the adiabatic energy potential is modified after photoexcitation to lead the minimum energy position to finite atomic displacements corresponding to the $A_{1g}$ phonon [52]. As a result, the $A_{1g}$ phonon is excited instantaneously and coherently.

To investigate the origin of the phase inversion, we conducted band structure calculations based on the DFT with the modulated crystal structures corresponding to the $A_{1g}$ phonon. From the calculation results, it was found that the hole FS mainly originated from the $d_{z^2}$ orbital strongly warped with a decrease in the pnictogen height. Specifically, the $d_{z^2}$ hole FS is strongly warped for the lower $h$ value whereas the warping is dramatically weakened for the higher $h$ value. By contrast, the modulation of the electron pockets is inverted with respect to that of the $d_{z^2}$ hole FS; that is, they become larger for higher $h$ and smaller for lower $h$. These opposite behaviors between the hole and electron FSs account for the observed antiphase oscillations. More importantly, we found that the pnictogen height decreases, and this direction is the same as that induced through the substitution of P for As, in which superconductivity is induced by a structural modification without carrier doping [50].

### 3.1.2 FeSe

FeSe has the simplest crystal structure among the ion-based superconductors. It is also noteworthy that it exhibits no magnetic order in contrast to the other iron-based superconductors. One of the significant aspects of FeSe is the dramatic increase in $T_c$ under various external stimuli. Physical pressure drives $T_c$ up to ~40 K [53] [54] [55], whereas the intercalation of the spacer layers can increase $T_c$ to ~40 K [56] [57]. These reports suggest that the electronic properties of FeSe can be easily manipulated, and we have regarded photoexcitation as an alternative control tool with many substantial advantages over other methods.

Figures 3(a) and 3(b) show the momentum-integrated TARPES spectra across the hole and electron FSs as a function of energy with respect to $E_F$ and pump-probe delay time. The integrated intensities above $E_F$ corresponding to the boxed regions in Figs. 3(a) and 3(b) are shown in Figs. 3(c) and 3(d), respectively. In contrast to BaFe$_2$As$_2$, an immediate excitation and overshooting decay at $\Delta t$ of ~0 ps followed by relatively slow relaxation dynamics at both FSs were observed. At the larger delay time of $\Delta t$ ~3.0 ps, whereas the intensity of the hole FS decreases, that of the electron FS increases, which will be further discussed later. Similar to BaFe$_2$As$_2$, the oscillatory behavior was clearly observed to be superimposed onto the background carrier dynamics. To highlight the oscillatory components, we subtracted the background carrier dynamics shown by the black solid lines shown in Figs.

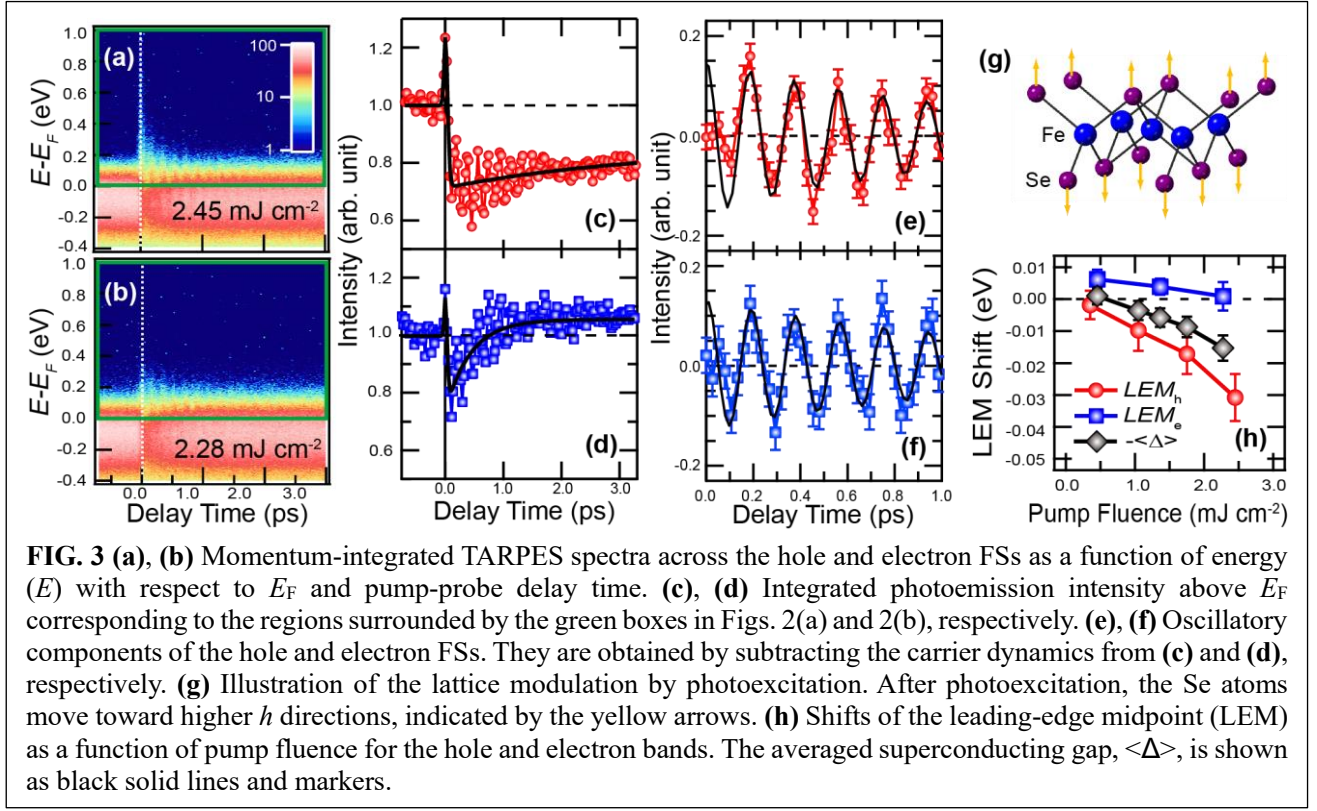

**FIG. 3 (a)**, **(b)** Momentum-integrated TARPES spectra across the hole and electron FSs as a function of energy ($E$) with respect to $E_F$ and pump-probe delay time. **(c)**, **(d)** Integrated photoemission intensity above $E_F$ corresponding to the regions surrounded by the green boxes in Figs. 2(a) and 2(b), respectively. **(e)**, **(f)** Oscillatory components of the hole and electron FSs. They are obtained by subtracting the carrier dynamics from **(c)** and **(d)**, respectively. **(g)** Illustration of the lattice modulation by photoexcitation. After photoexcitation, the Se atoms move toward higher $h$ directions, indicated by the yellow arrows. **(h)** Shifts of the leading-edge midpoint (LEM) as a function of pump fluence for the hole and electron bands. The averaged superconducting gap, <Δ>, is shown as black solid lines and markers.

3(c) and 3(d). They are shown in Figs. 3(e) and 3(f) with the fitting of the damped oscillation functions. From these results, we found that the oscillations are cosine-like with a frequency of 5.3 THz and are in phase with respect to each other, which is a stark difference from the oscillations observed in BaFe$_2$As$_2$ [26]. From the comparison with the Reman result [58], this oscillation frequency is assigned to the $A_{1g}$ phonon mode, in which two Se layers oscillate symmetrically with respect to the sandwiched Fe layer. The cosine-like feature also confirms that the observed oscillation is based on the DECP mechanisms. From the comparison with the DFT calculations as conducted for BaFe$_2$As$_2$, we also found that the new stable (metastable) states have higher Se heights measured from the nearest Fe layer, $h$, compared to the equilibrium state, as indicated by the yellow arrows in Fig. 3(g). Interestingly, the realized lattice modulations were opposite those of BaFe$_2$As$_2$, in which $h$ becomes lower after photoexcitation.

As we mentioned that it will be further discussed for the contrasting behavior between the hole and electron FSs at the relatively large delay time of $\Delta t \sim 3.0$ ps, we proceeded to measure the long-delay time behaviors of the TARPES spectra for both hole and electron FSs until ~1 ns. The leading-edge midpoint (LEM) shifts for the hole and electron FSs under long delay times are shown as blue and red markers in Fig. 3(h), respectively, as a function of the pump fluence. Based on the carrier conservation, photoexcited electrons from the hole bands are considered to be relaxed to the electron bands. Thus, the amount of LEM shifts of the hole FS is expected to be comparable and has a sign opposite that of the electron FS. However, the LEM shift of the electron FS decreases with an increase in the pump fluence, as shown in Fig. 3(f). This unusual behavior can be explained by considering the overall LEM shifts. After considering and excluding all other possible effects such as surface photovoltage effect, multiphoton effect, and Floquet states, the overall shift can be ascribed to a superconducting-like state characterized by the gap, Δ, which is plotted as the black markers in Fig. 3(h). The possibility of the superconducting state is also supported by the observed lattice change of a higher $h$, through which increasing $T_c$ has been confirmed under physical pressure [55].

### 3.2 Graphene

Owing to its unique physical, electronic, and chemical properties, numerous investigations have been conducted on a carbon-sheet material, *graphene* [59]. Optical properties have also attracted significant attention, and many singular phenomena have been reported, such as multiple carrier generations [60] [61] [62] [63] or phonon bottleneck effects [64] [65] [66]. These phenomena are determined by the dynamics of fermions in a linearly dispersed band structure, that is, a Dirac cone. This massless band structure can be modified into a massive structure by introducing another sheet attached [67], called *bilayer graphene*. As a result, the carrier dynamics in bilayer graphene show a different behavior from that of single-layer graphene [68]. Furthermore, the introduction of the twisting angle

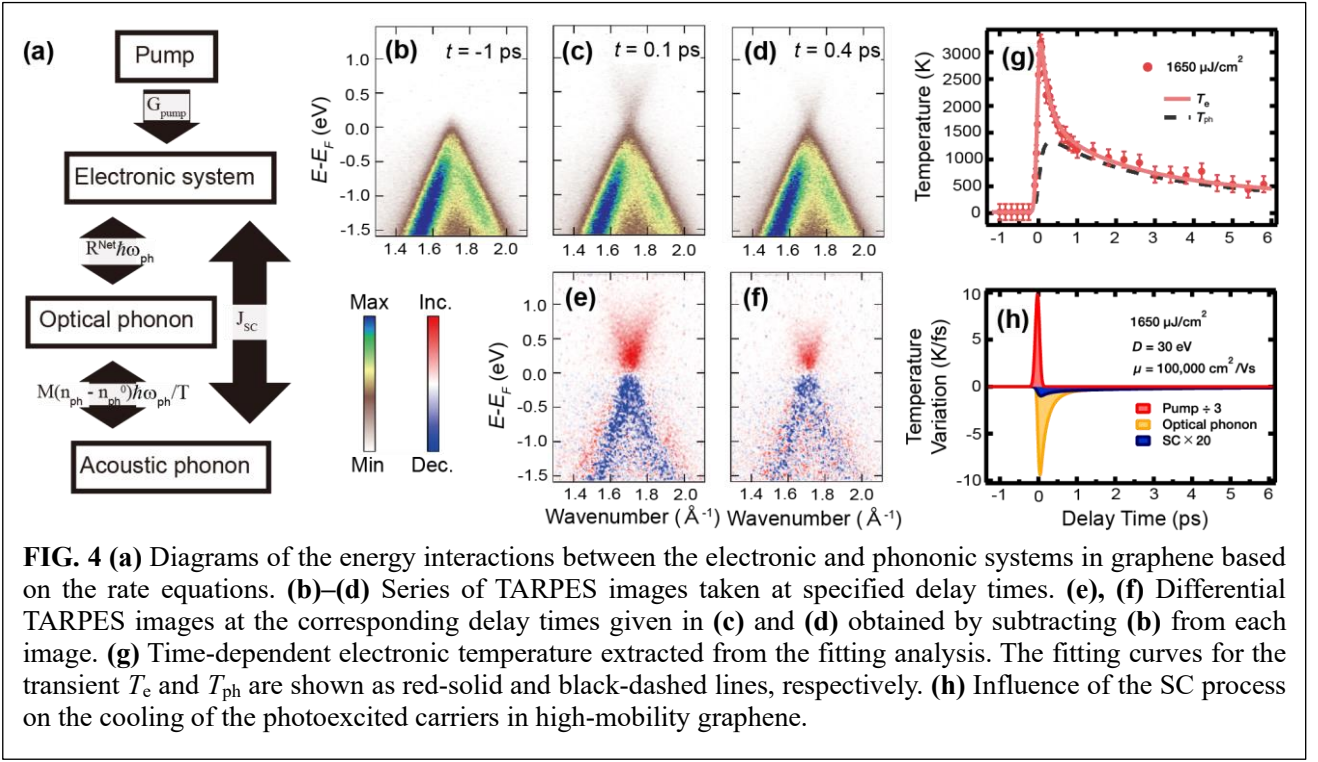

**FIG. 4 (a)** Diagrams of the energy interactions between the electronic and phononic systems in graphene based on the rate equations. **(b)–(d)** Series of TARPES images taken at specified delay times. **(e), (f)** Differential TARPES images at the corresponding delay times given in (c) and (d) obtained by subtracting (b) from each image. **(g)** Time-dependent electronic temperature extracted from the fitting analysis. The fitting curves for the transient $T_e$ and $T_{ph}$ are shown as red-solid and black-dashed lines, respectively. **(h)** Influence of the SC process on the cooling of the photoexcited carriers in high-mobility graphene.

between each layer in the bilayer graphene can dramatically modify the electronic properties [69] [70] [71] [72] [73].

In this respect, TARPES is an extremely suitable tool because it can directly track the dynamic evolution of electrons in the band dispersion after photoexcitation. Moreover, the relatively high photon energy obtained by HHG is necessary to access the Dirac cone, which lies at the boundary of the Brillouin zone. We conducted HHG laser TARPES to observe the carrier dynamics in high-mobility graphene [28] and quasi-crystalline twisted bilayer graphene [29]. The carrier dynamics were found to be highly sensitive to the layer structures.

### 3.2.1 High-mobility graphene

It is essential to study the carrier dynamics of graphene for applications in optoelectronic devices as well as fundamental interest. In particular, a comprehensive understanding of the carrier cooling dynamics in photoexcited graphene is desired. Typically, the scattering of electrons by acoustic phonons can transfer less energy compared to optical phonons. Introducing an impurity that promotes carrier cooling through three-body scattering among carriers, acoustic phonons, and impurities, which are called super collisions (SCs) [74] and are schematically shown in Fig. 4(a), can be considered to lower the efficiency of energy harvesting devices. However, SCs have yet to be clarified owing to the complicated interplay between carriers, optical phonons, acoustic phonons, and defects. In this study, we conducted TARPES measurements on graphene grown on a SiC(000$\bar{1}$) C-terminated surface, for which the intrinsic carrier mobility exceeded 100,000 cm$^2$V$^{-1}$s$^{-1}$, and conducted simulations based on a two-temperature model to study the contribution of SCs as a cooling channel.

Figures 4(b)–4(d) show the TARPES image at delay times of −1.0, 0.1, and 0.4 ps. To highlight the pump-induced changes, the differential images are also shown in Figs. 4(e) and 4(f) by subtracting the image before the arrival of the pump from each image. The electrons are immediately transferred from the occupied lower Dirac cone to the unoccupied upper cone and relaxes to the original state. From the fitting analysis, temporal electron temperature is plotted as red markers in Fig. 4(g).

To understand the underlying relaxation processes, we investigated the energetic interchanges between the electronic and phononic systems, as well as energy dissipation through SCs. The diagrams of these energy interactions are shown in Fig. 4(a). Here, $G$ represents the injected energy into the sample during the laser irradiation while $R$ denotes the net recombination rate. $M$ denotes the number of phonon modes for the carrier-phonon scattering, and $J_{sc}$ is the energy loss rates of SC. We considered intravalley ($\hbar\omega_{ph} = 196$ meV) and intervalley ($\hbar\omega_{ph} = 160$ meV) optical phonons for the scatterings with electrons. Time-dependent electronic and optical phonon temperatures reproduced by solving a set of rate equations based on the two-temperature model are shown as red-solid and black-dashed lines, respectively, in Fig. 4(g). Figure 4(h) shows the term-by-term comparisons of the calculation results for $dT_e/dt$; specifically, the heating/cooling rates using a pumping laser, optical phonons, and SCs are displayed separately. Here, $D$ and $\mu$ are the deformation potential and intrinsic carrier mobility, respectively. The

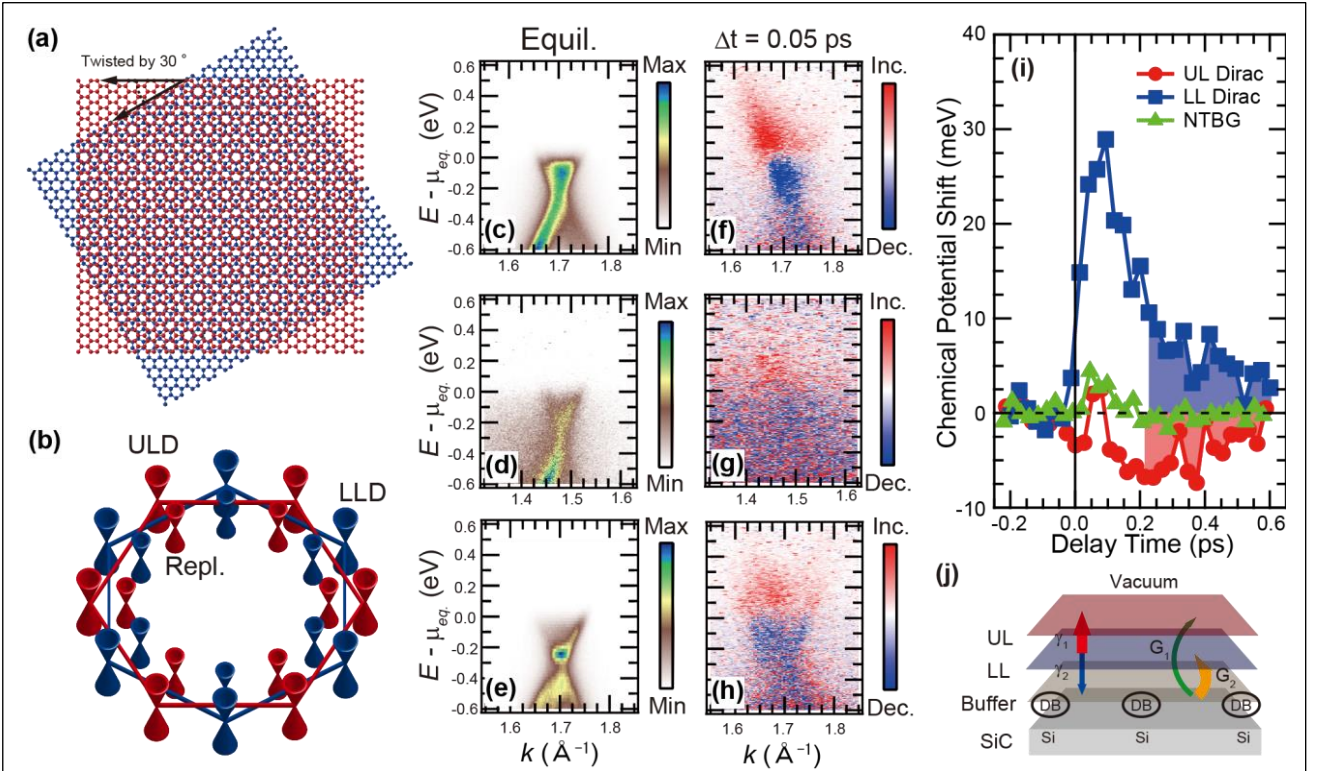

**FIG. 5 (a)** Crystal structure of quasicrystalline 30 twisted bilayer graphene (QCTBG). **(b)** Schematic drawing of the electronic structures of QCTBG in momentum space. Outer red and blue Dirac cones represent the upper-layer Dirac (ULD) and lower-layer Dirac (LLD) bands, respectively, whereas the inner red and blue Dirac cones correspond to the replica bands of the ULD and LLD bands. **(c)–(e)** ARPES image for the ULD, LLD, and NTBG bands in equilibrium. **(f)–(h)** Differential TARPES images for the ULD, LLD, and NTBG bands. Red and blue points represent increasing and decreasing photoemission intensities, respectively. **(i)** Chemical-potential shift as a function of pump-probe delay time for the ULD and LLD bands. For comparison, the result for NTBG is also shown. **(j)** Schematic illustration of the spatial relationship among the upper layer (UL), lower layer (LL), buffer layer, and SiC substrate in QCTBG. Schematic illustrations of the carrier transport among the layers in QCTBG are shown by arrows with parameters. The thicker lines indicate that $\gamma_1$ and $G_2$ are larger than $\gamma_2$ and $G_1$, respectively.

expression for $J_{SC}$ takes the form

$$J_{SC} \sim 8.8 \times 10^{14} \times \frac{D^2}{\mu}(T_e^3 - T_{ac}^3),$$

where $T_{ac}$ is the acoustic phonon temperature, which is assumed to be unchanged from the equilibrium state. Comparing the cooling power through the SCs (integral of the blue area) with the total cooling power (sum of integral of the yellow and blue areas), the SCs contribute to carrier cooling based on a ratio of 1.1%, from which SCs are found to have a negligible influence on decreasing the electronic temperature in extremely high-mobility graphene. This finding in the present C-faced graphene is in contrast to the case of Si-faced graphene, where SCs much more frequently occur and have more dominant role in the cooling process [66]. Our findings provide clear guidelines for designing next-generation optoelectronic devices and improving their performance.

### 3.2.2 Quasicrystalline twisted bilayer graphene

Twisted bilayer graphene has led to many exotic quantum phenomena [69] [70] [71] [72] [73]. The twist angle $\Theta$ has recently become known as an important degree of freedom for realizing a variety of exotic states of this material; that is, the Mott insulating state and two-dimensional superconducting state ($\Theta = 1.1°$). At a twisting angle of $\Theta = 30°$, the crystal structure acquires quasi-crystallinity, where translational symmetry is absent, as shown in Fig. 5 (a) [75] [76]. As a result, the electronic structure is significantly affected, and the band structure exhibits interesting features. Figure 5(b) shows a schematic illustration of the electronic band structure of quasi-crystalline twisted bilayer graphene (QCTBG). The outer red and blue Dirac cones represent the upper layer Dirac (ULD) and lower layer Dirac (LLD) cones, respectively. As a result of the strong interlayer interaction connected by the Umklapp scattering in each layer, the replica bands of the ULD and LLD are created, as shown in the inner red and blue Dirac cones in Fig. 5(b). To understand the nonequilibrium properties and evaluate the potential for

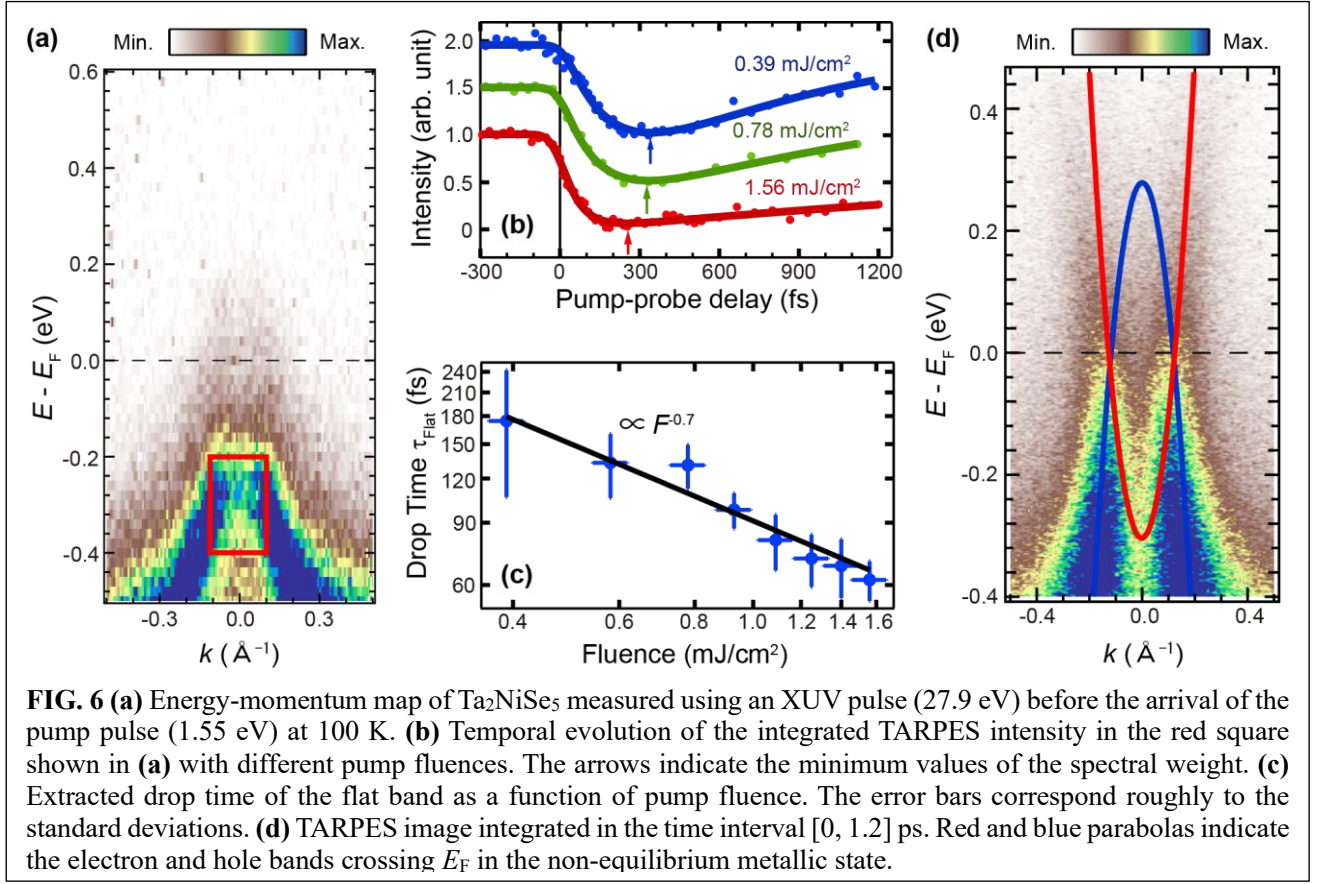

**FIG. 6 (a)** Energy-momentum map of $Ta_2NiSe_5$ measured using an XUV pulse (27.9 eV) before the arrival of the pump pulse (1.55 eV) at 100 K. **(b)** Temporal evolution of the integrated TARPES intensity in the red square shown in **(a)** with different pump fluences. The arrows indicate the minimum values of the spectral weight. **(c)** Extracted drop time of the flat band as a function of pump fluence. The error bars correspond roughly to the standard deviations. **(d)** TARPES image integrated in the time interval [0, 1.2] ps. Red and blue parabolas indicate the electron and hole bands crossing $E_F$ in the non-equilibrium metallic state.

application, it is essential to study the ultrafast carrier dynamics in this system. To this end, we studied the ultrafast dynamics of QCTBG by comparing the results for non-twisted bilayer graphene (NTBG) as a reference [29].

Figures 5(c)–5(e) show the equilibrium ARPES images for the ULD, LLD, and NTBG bands, respectively. The Dirac cones of QCTBG are *n*-type, where the Dirac points are located below the equilibrium chemical potential ($\mu_{eq}$). The electronic structure of NTBG is also an *n*-type, and there is a band gap at 0.3 eV below $\mu_{eq}$. After the pump pulse of 0.7 mJ/cm$^2$, the TARPES band diagram of the individual bands of bilayers evolves along the femtosecond time scale. To enhance the temporal variations, differential TARPES images are shown as a difference between the images before and after photoexcitation, where the red and blue regions in Figs. 5(f)–5(h) correspond to the increase and decrease of the photoemission intensity at the delay time ($\Delta t$) of 0.05 ps, respectively. The spectral weights of the bands below $\mu_{eq}$ decrease and those above $\mu_{eq}$ immediately increase. This reflects the excitation of electrons from the occupied bands to the unoccupied bands.

To evaluate the occupation of the Dirac cones by nonequilibrium carriers, we fit the energy distribution curves of each TARPES image using the Fermi-Dirac distribution function convoluted using a Gaussian, which extracts the electronic temperature and chemical potential shift. Figure 5(i) shows the temporal chemical potential shifts for the ULD, LLD, and NTBG. Surprisingly, the opposite behavior between the ULD and LLD was observed, that is, ULD underwent a negative shift, whereas LLD underwent a positive shift. It was also found that the chemical potential shift for the NTBG band remained constant at essentially zero over the delay time from 0.1 to 0.6 ps. The striking difference among the three types of Dirac cones provides clear evidence of a carrier imbalance between the ULD and LLD bands of QCTBG at the ultrafast time scale.

To understand the underlying mechanism, we conducted calculations based on the rate equations shown as follows.

$$\frac{dn_{el}^{UL}}{dt} = -\frac{n_{el}^{UL}}{\tau_{UL}} + \gamma_1(n_{el}^{LL} - n_{el}^{UL}) + G_1 exp\left(-\frac{t^2}{T_p^2}\right),$$

$$\frac{dn_{el}^{LL}}{dt} = -\frac{n_{el}^{LL}}{\tau_{LL}} - \gamma_1(n_{el}^{LL} - n_{el}^{UL}) - \gamma_2(n_{el}^{LL} - n_{el}^{Sub})$$

$$+ G_2 exp\left(-\frac{t^2}{T_p^2}\right),$$

$$n_{el}^{UL} + n_{el}^{LL} + n_{el}^{Sub} = const.,$$

where $n_{el}^{UL/LL}$ and $\tau_{UL/LL}$ are electron densities and lifetime in the ULD/LLD band of a QCTBG, respectively. $\gamma_1$ and $\gamma_2$ are rate constants of arrier transfer between the UL and LL, and the LL and SiC substrate. $G_1$ and $G_2$ are coefficients of pump induced net density flux to the UL and LL from SiC substrate, respectively. $T_p$ is the time width, reflecting the temporal resolution. $n_{el}^{Sub}$ is the electron density in the SiC substrate.

The spatial relation between the parameters for the upper layer (UL), lower layer (LL), and SiC substrate used in the equations is shown in Fig. 5(j). The carrier transfer rates deduced from the calculations are indicated by the width of the arrows in Fig. 5(j). We found that $\gamma_1$ of larger than $\gamma_2$ indicates that the carrier transfer is more frequent between the graphene layers than between the LL and the substrate. Furthermore, the finite values of $G_1$ and $G_2$ demonstrate that transient carrier doping from the substrate to each graphene layer exists, and that a $G_2$ of larger than $G_1$ is the origin of the observed unbalanced carrier distribution between the UL and LL. The present investigations demonstrate the feasibility of manipulating the dynamics of Dirac carriers in individual layers of bilayer graphene and provide valuable information for designing future graphene-based ultrafast optoelectronic devices.

### 3.3 Ta$_2$NiSe$_5$

A photoinduced phase transition is expected to be a key mechanism for next-generation devices because it can instantly change the properties of a material [77] [78] [79]. The realized state can be qualitatively different from the high-temperature phase in equilibrium with a higher entropy. The underlying mechanisms of such phenomena are intertwined interactions between the charge, spin, and lattice degrees of freedom [80]. In this respect, strongly correlated electron systems provide extremely attractive playgrounds for various photoinduced phase transitions because they exhibit rich phase diagrams owing to the subtle balance among competing orders in equilibrium [81], and can be relatively easily manipulated by external stimuli such as the physical pressure [82] or magnetic field [83].

In this subsection, we review our recently studied material, Ta$_2$NiSe$_5$, which is regarded as a unique candidate for an excitonic insulator [84]. We found that the response time measured using TARPES on Ta$_2$NiSe$_5$ reveals the characteristics of an excitonic insulator. Furthermore, we discovered a photo-induced metallic phase in Ta$_2$NiSe$_5$, which was also confirmed to be different from the high-temperature phase in equilibrium [30]. To investigate the photo-induced insulator-to-metal transition in terms of electron-phonon couplings, we developed a novel analysis method called frequency-domain ARPES (FDARPES). This method can reveal the underlying nature of photo-induced phase transitions through the electron–phonon coupling [31].

#### 3.3.1 Photo-induced semimetallic state

Figure 6(a) shows an energy-momentum (*E-k*) TARPES intensity map of Ta$_2$NiSe$_5$ around the Γ point (center of the Brillouin zone) taken before the arrival of the pump pulse at 100 K. To visualize how the flat band, which has been considered a characteristic of an excitonic insulator, collapses after pump excitation, we show the temporal evolution of the integrated TARPES intensity in Fig. 6(b) for several pump fluences. The rectangular region shown in Fig. 6(a) shows the integration range. It can be observed that the initial decrease in the TARPES intensity depends strongly on the pump fluence and becomes faster with increasing pump fluence. To evaluate the drop time of the flat band ($\tau_{Flat}$), which is the time scale of the intensity decrease of the flat band after pumping, the data were fitted to a Gaussian-convoluted rise-and-decay function, and the obtained values of $\tau_{Flat}$ are plotted as blue symbols in Fig. 6(c).

The time scale of the gap collapse in excitonic insulators is considered to be inversely proportional to the plasma frequency, $\omega_p = (ne^2/\varepsilon_0\varepsilon_r m^*)^{1/2}$, where $n$ is the carrier density, $e$ is the elementary charge, $m^*$ is the effective mass of the valence or conduction band, $\varepsilon_0$ is the electric constant, and $\varepsilon_r$ is the dielectric constant. From this relationship, the gap quenching time should be proportional to $n^{-1/2}$. We found that the drop time was proportional to $F^{-0.7}$, where $F$ is the pump fluence. The deviation from the ideal value of 0.5 is considered to be due to non-negligible nonlinear absorptions, such as two or three photon absorptions. Similar photo-excitation behavior has been observed for 1*T*-TiSe$_2$, which has been considered as another candidate excitonic insulator [85]. This finding strongly suggests that Ta$_2$NiSe$_5$ is an excitonic insulator.

Next, we show a more impressive temporal behavior in the TARPES image of Ta$_2$NiSe$_5$. Figure 6(d) shows a time-integrated TARPES image after pumping. Both the electron and hole bands cross $E_F$ at the same Fermi momenta of $k_F \sim \pm 0.1$ Å$^{-1}$, as schematically shown by the red and blue parabolas in Fig. 6(d). This may indicate that the hybridization between the two Ta chains is sufficiently strong to lift the degeneracy. However, because this is not predicted by band-structure calculations, the emergence of the hole and electron bands crossing $E_F$ at the same $k_F$ is a surprising nature of the observed non-equilibrium metallic phase, which indicates that the observed non-equilibrium metallic state is entirely different from the high-temperature phase in a state of equilibrium.

#### 3.3.2 Frequency-domain ARPES

To investigate the photo-induced insulator-to-metal transition observed in Ta$_2$NiSe$_5$ from the viewpoint of electron-phonon interactions, we used a novel analysis method called frequency-domain ARPES (FDARPES), to which similar analysis has been performed by Hein *et al.* [86]. Figure 7(a) shows the differential TARPES images before and after pump excitation. The peak positions in the TARPES images are indicated by the circles in Fig. 7(a). To reveal the photo-induced profile in Ta$_2$NiSe$_5$ more specifically, we investigated the TARPES images in terms of the electron–phonon couplings. First, we analyzed the time-dependent intensities for the representative energy and momentum ($E$-$k$) regions, indicated as I–IV in Fig. 7(a). The data are composed of the background carrier dynamics superimposed by the oscillations, which result from the excitation of coherent phonons. To extract the oscillatory components, we first fit the carrier dynamics to a double-exponential function convoluted with a Gaussian, and then subtract the fitting curves from the data. Fourier transforms are performed for the subtracted data, and the intensities for each $E$-$k$ region are shown in Fig. 7(b). The peak structures appear distinctively depending on the $E$-$k$ regions.

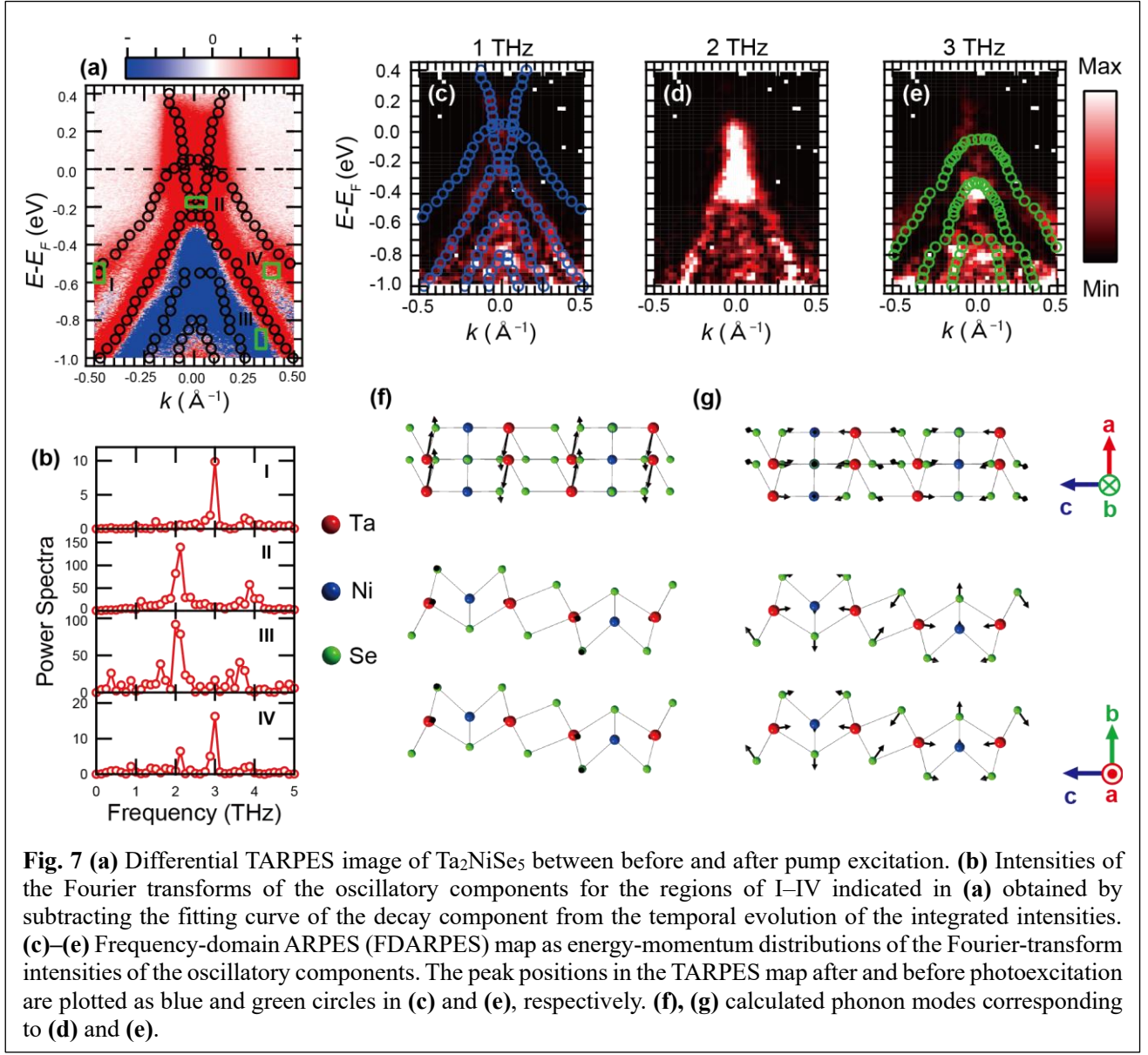

**Fig. 7** **(a)** Differential TARPES image of Ta$_2$NiSe$_5$ between before and after pump excitation. **(b)** Intensities of the Fourier transforms of the oscillatory components for the regions of I–IV indicated in **(a)** obtained by subtracting the fitting curve of the decay component from the temporal evolution of the integrated intensities. **(c)–(e)** Frequency-domain ARPES (FDARPES) map as energy-momentum distributions of the Fourier-transform intensities of the oscillatory components. The peak positions in the TARPES map after and before photoexcitation are plotted as blue and green circles in **(c)** and **(e)**, respectively. **(f), (g)** calculated phonon modes corresponding to **(d)** and **(e)**.

To investigate the electron–phonon couplings in more detail, we further mapped out the peak intensity of the Fourier component as a function of energy and momentum for each peak frequency, which we call the FDARPES map. Figures 7(c)–7(e) show the FDARPES maps corresponding to frequencies of 1, 2, and 3 THz, respectively. To observe each phonon mode associated with the FDARPES map, we conducted *ab initio* calculations. The calculated phonon modes corresponding to 2 and 3 THz are shown in Figs. 7(f) and 7(g), respectively. Noticeably, the FDARPES map exhibits significantly different behaviors depending on the frequency, which demonstrates that each phonon mode is selectively coupled to specific electronic bands. In particular, the 2-THz phonon mode has the strongest signal near $E_F$. According to a recent theoretical investigation which reported that the intensity of FDARPES spectra is expressed as the sum of two terms proportional to diagonal and off-diagonal electron-phonon coupling matrix elements when the phonon induced variation in band energy is small enough [87], this strongest signal can be ascribed to the situation, where the 2-THz phonon mode is most strongly coupled to the emergent photoinduced electronic bands crossing $E_F$.

To highlight the spectral features of each FDARPES map in more detail, we compared the FDARPES map with the band dispersions deduced from the peak positions of the TARPES images before and after photoexcitation. We found that the FDARPES map of 1 THz matches the band dispersions after photoexcitation better than those before photoexcitation, as shown in Fig. 7(c), whereas the FDARPES map of 3 THz is closer to the band dispersions before photoexcitation, as shown in Fig. 7(e). This means that both semimetallic and semiconducting bands coexist in the transient state, suggesting that the strong electron–phonon couplings for the 1- and 3-THz phonon modes are associated with the semimetallic and semiconducting bands, respectively. Moreover, this finding demonstrates that, whereas the semimetallic and semiconducting states coexist after photoexcitation, the FDARPES method can selectively detect the coupling of each phonon mode to semimetallic and semiconducting states in a frequency-resolved manner.

4  Summary

We briefly reviewed our recent study conducted using HHG laser TARPES. The nonequilibrium profiles of various quantum materials were revealed through observations of their temporal electronic band structures. To induce further exotic profiles, it is desirable to extend the wavelength of the pump pulses to the regions of the resonant excitations of infrared phonons or band gaps of semiconductors. For the exotic states induced through light illumination, many unexplored phenomena are awaiting to be revealed, such as photoinduced topological phase transitions, in addition to photoinduced insulator-to-metal transitions, and photoinduced superconductivity. The recent advances of the terahertz technology enable us to obtain strong sub cycle field exceeding 1 MV/cm [88]. By using such a sub cycle pulse as a pump, we can explore field-induced phenomena in the whole Brillouin zone. Furthermore, the improvement of time resolution in the range of attosecond regime can reveal many fundamental phenomena [89]. By implementing as a pump-probe geometry, we can explore initial dynamics for the excited or unoccupied states.

Acknowledgements

We would like to thank Editage (www.editage.com) for the English language editing. This work was supported by Grants-in-Aid for Scientific Research (KAKENHI) (Grant Nos. JP18K13498, JP19H01818, and JP19H00651) from the Japan Society for the Promotion of Science (JSPS), by JSPS KAKENHI on Innovative Areas, "Quantum Liquid Crystals" (Grant No. JP19H05826), and by the MEXT Quantum Leap Flagship Program (MEXT Q-LEAP) (Grant No. JPMXS0118068681), Japan

*Corresponding author.
takeshi.suzuki@issp.u-tokyo.ac.jp

†Corresponding author.
okazaki@issp.u-tokyo.ac.jp